\documentclass[aps,showpacs,twocolumn,superscriptaddress]{revtex4}
\usepackage{graphicx}
\begin{document}
\title{Wobbling excitations at high spins in  A$\sim$160 }
\author{J. Kvasil}
\affiliation{Institute of Particle and Nuclear Physics, Charles
University, V.Hole\v sovi\v ck\'ach 2, CZ-18000 Praha 8, Czech Republic}
\author{R. G. Nazmitdinov}
\affiliation{Departament de F{\'\i}sica,
Universitat de les Illes Balears, E-07122 Palma de Mallorca, Spain}
\affiliation{Bogoliubov Laboratory of Theoretical Physics,
Joint Institute for Nuclear Research, 141980 Dubna, Russia}

\date{\today}
\begin{abstract}
We found that in $^{156}$Dy and $^{162}$Yb the lowest odd spin gamma-vibrational states  
transform to the wobbling excitations after the 
backbending, associated with the transition from axially-symmetric to nonaxial 
shapes.  The analysis of quadrupole electric transitions
determines uniquely the sign of the $\gamma$-deformation in  
both nuclei after the transition point.
\end{abstract}
\maketitle

Thanks to novel experimental detectors, a new frontier of discrete-line $\gamma$-spectroscopy
at very high spins has been opened in the rare-earth nuclei (see, for example, \cite{prl}).
These nuclei can accommodate the highest values of the angular
momentum, providing one with various nuclear structure phenomena.
The quest for manifestations of nonaxial deformation is one of the driving forces
in high spin physics in past few years \cite{fra}. The identification of wobbling excitations is 
recognized nowadays as a convincing proof of the nonaxiality.
Wobbling excitations were suggested first by Bohr and Mottelson for rotating 
even-even nuclei \cite{BM75}
and studied soon within simplified microscopic models \cite{mic} 
(see also Ref.\onlinecite{shi1} and references therein). According to 
the microscopic approach \cite{JM79,mar2}, the wobbling excitations
are  vibrational states of the negative signature
built on the positive signature yrast (vacuum) state. Their characteristic feature
is collective E2 transitions with $\Delta I=\pm 1 \hbar$ between these and yrast states.
First experimental evidence of such states in odd Lu nuclei was reported
only recently \cite{OH01}. 

The first analysis of the properties  of the second 
triaxial superdeformed band in $^{163}$Lu  was based upon phenomenological 
particle-rotor calculations \cite{ham}. The absolute values of the 
irrotational moments of inertia were fitted and  
so-called "$\gamma$-reversed" dependence of these moments was introduced in order to
obtain a reasonable agreement with the experimental data. 
It was shown in Ref.\onlinecite{mat} that the microscopic approach \cite{mar2} 
may gain a better insight into the observed phenomena.
In the analysis of \cite{mat}, however, the constant mean-field deformation parameters 
are used, which is not always justified.
Moreover,  the authors admitted that the kinematic 
moment of inertia $\Im_x$ was not described properly due to
the strong velocity dependence of the Nilsson potential
(see discussion in Ref.\onlinecite{mat}). 
We recall that wobbling excitations depend on  all three moments of inertia that
characterize the nonaxial shape. Therefore, a self-consistent 
description of  moments of inertia is a prerequisite of the  microscopic
analysis of the nuclear wobbling motion.
The main aim of this Letter is to analyze  new  data on high spin
states in $^{156}$Dy and $^{162}$Yb \cite{nndc,Kon}  within a 
microscopic approach \cite{JR} based on  the cranked
Nilsson model plus random phase approximation (CRPA).
In our approach mean field parameters are determined from the  
energy-minimization procedure. 
The proper description of the moment inertia $\Im_x$ is achieved 
using the recipe suggested in Ref.\onlinecite{nak}.
Our calculations suggest that some excited states at high spins
may represent wobbling excitations.

Our model Hamiltonian is
\begin{equation}
\hat H_{\Omega} \,= \hat H_0 -\sum_{\tau} \lambda_{\tau} \hat N_{\tau}
-\Omega \hat J_x+V
\label{h1}
\end{equation}
The term  $\hat H_0=\hat H_N\,+ \hat H_{\rm add}$ contains the Nilsson Hamiltonian
$\hat H_N$ and the additional term that restores the local Galilean invariance
of the Nilsson potential, broken in the rotating frame \cite{nak}.
This term is essential to obtain a correct description of 
$\Im_x$-moment of inertia \cite{JR}. 
Although the additional term $\hat H_{\rm add}$ breaks the rotational symmetry
in the sense of Eq.(\ref{spur}) (see below), this effect can be negligibly small 
in the RPA order. The chemical potentials $\lambda_\tau$ $(\tau=$n or p) are
determined so as to give correct average particle numbers $\langle \hat N_\tau \rangle$.
Hereafter, $\langle...\rangle$ means the averaging over the mean field vacuum (yrast)
state at a given rotational frequency $\Omega$.
The interaction $V$ includes separable monopole pairing,  monopole-monopole,
and quadrupole-quadrupole terms to describe the positive parity states.
All multipole and spin-multipole operators have a good isospin $T$ and signature
$r=\pm 1$ (see the properties of the matrix elements in Ref.\onlinecite{KV}).
They  are expressed in terms of doubly stretched coordinates
$\tilde x_i = (\omega_i/\omega_0)\,x_i$, which ensure the self-consistent conditions
at the equilibrium deformation. Details about the model Hamiltonian (\ref{h1}) can be found in
Refs.\onlinecite{JR}. 

The Nilsson-Strutinsky analysis of experimental data 
on high spins in $^{156}$Dy \cite{Kon} indicates that the positive parity yrast 
sequence undergoes a transition from the prolate  towards the oblate 
rotation. In our calculations the deformation parameters $\beta$ and $\gamma$ are  
defined by means of the oscillator frequencies 
$\omega^{2}_{i} = \omega^{2}_{0}\left[ 1 - 2\beta \sqrt{\frac{5}{4\pi}}cos(\gamma - \frac{2\pi}{3}i)\right]$ 
$( i = 1,2,3$ or $x,y,z)$. To compare our results with available experimental data \cite{Kon}, 
we consider the mesh on the ${\beta, \gamma}$ plane: from $\gamma=60^0$ (an oblate rotation around 
the y-axis) to $\gamma=-60^0$ (an oblate rotation around the x-axis) and $\beta=0-0.6$. 
At each rotational frequency, we have determined the equilibrium deformation 
parameters $(\beta,\gamma)$ by minimizing the mean-field energy 
$E_{MF}=\langle \hat H_{\Omega}\rangle$ on the mesh. 
In the vicinity of the backbending this procedure becomes highly unstable. In order 
to avoid unwanted singularities for certain values of $\Omega$, we followed the 
phenomenological prescription \cite{wys} for the definition of the pairing gap parameter
(see details in Refs.\onlinecite{JR}).  Parameters of the Nilsson potential
were taken from Ref.\cite{shel}. In our calculations we include all shells up to
$N=9$. Near the transition point we extended our
configuration space up to $N=10$ shells. The difference between results from the
former and the latter cases was small and all presented results are obtained with
$N=0-9$ shells. In contrast to  standard calculations with the Nilsson potential,
based on a "single stretched" coordinate method (cf \cite{Nil}), we 
use the real (non stretched) $ls$ and $l^2$ potentials, taking into complete account
$\Delta N = 2$ mixing produced by them. This improves the accuracy of  the mean
field calculations, since the "single stretched" $ls$ and $l^2$ potentials break
the rotational symmetry.

\begin{figure}[ht]
\includegraphics[height=0.22\textheight,clip]{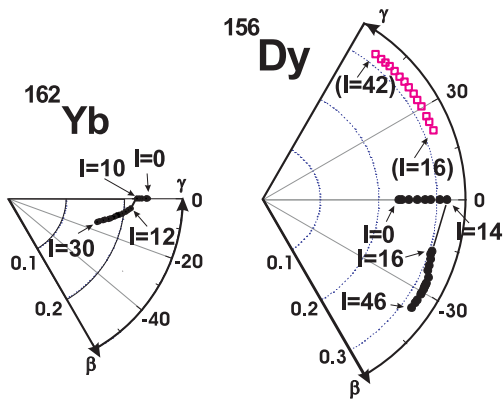}
\caption{(Color online)
Equilibrium deformations in $\beta$-$\gamma$ plane
as a function of the angular momentum $I = \langle {\hat J}_x \rangle-1/2$
(in units of $\hbar$). The equilibrium deformations  for $^{156}$Dy provide the 
lower mean field energies in the region $-\pi/3<\gamma<0$ (filled circles)  
in comparison with those (open squares) obtained in Ref.\onlinecite{JR}. 
The maximal difference between the minimal energies 
at the positive and negative equilibrium $\gamma$-values does not exceed $\sim 1$MeV 
for $^{156}$Dy.}
\label{fig1}
\end{figure}

Our results conform to the results of the 
Nilsson+Strutinsky shell correction method (compare our Fig.1 with Fig.3c 
in Ref.\onlinecite{Kon}), although we obtain slightly different values for the 
equilibrium deformations. In the analysis of Ref.\onlinecite{Kon} the pairing 
correlations are missing, while the hexadecapole deformation is not included in 
the present calculations. The triaxiality of the mean field sets in at
the critical rotational frequency $\hbar\Omega_c$ 
which triggers the backbending in the considered nuclei 
due to different mechanisms. We obtain  $\hbar \Omega_{c}\approx 0.25$MeV 
$(10\hbar\rightarrow 12\hbar)$ and $\hbar \Omega_{c}\approx0.3$MeV 
$(14\hbar\rightarrow 16\hbar)$ for $^{162}$Yb and $^{156}$Dy, respectively.
The contribution of the additional term was crucial to achieve a good correspondence 
between the calculated and experimental values of the crossing frequency in each nucleus.
In $^{156}$Dy  we obtain that the $\gamma$-vibrational excitation ($K=2$)
of the positive signature tends to a zero in the rotating frame at the transition point,
in close agreement with experimental data. At the transition point
there are two indistinguishable mean-field energy minima with different
shapes: axially symmetric and strongly nonaxial. The increase of the rotational
frequency changes the axial shape to the nonaxial one with a negative $\gamma$-deformation
($\gamma \sim -20^o$).  In contrast, 
the axially symmetric configuration in $^{162}$Yb is replaced by the two-quasiparticle
one with a small negative $\gamma$-deformation. There, the backbending occurs due to
the rotational alignment of a neutron $i_{13/2}$ quasiparticle pair. The nonaxiality evolves quite
smoothly.

In the CRPA approach the positive $(r=+1)$ and negative $(r=-1)$ signature boson spaces are
not mixed, since the corresponding operators commute and
$H_{\Omega}=H_\Omega(r=+1) + H_\Omega(r=-1)$. 
The self-consistency between the mean field and the RPA calculations
is achieved by varying the strength constants of the pairing and multipole interactions 
in the RPA. It results in the separation of collective excitations  from those, related
to the symmetries broken by the mean field. 
Two zero solutions are associated with
the violation of the particle number (for protons and for neutrons) 
$\left[\hat H_\Omega(r=+) \,,\hat N_{\tau}\right] = 0$. 
The other one is related to the  spherical symmetries
of the mean field $\left[\hat H_\Omega(r=+)\,,\,\hat J_x \right] =0$.
While the positive signature excitations are analyzed in Ref.\onlinecite{JR},
the main focus of this Letter is wobbling excitations that belong to the negative
signature sector. The negative signature RPA Hamiltonian 
has the form 
\begin{equation}
\hat H_{\Omega}[r =-1]=\frac{1}{2}\sum_{\mu}E_{\mu}b^{+}_{\mu}b_{\mu}-
\frac{\chi}{2}\sum_{\mu_{3}=1,2}\tilde{Q}^{(-)2}_{\mu_3}\, ,
\label{hns}
\end{equation}
where $E_{\mu} = \varepsilon_{i} + \varepsilon_{j}$
($E_{\bar i \bar j} = \varepsilon_{\bar i} + \varepsilon_{\bar j}$)
are two-quasiparticle energies and $b^{+}_{\mu}(b_{\mu})$ is
a quasi-boson creation (annihilation) operator \cite{JR}.
Hereafter, 
the index $\mu$ runs over $ij$, $\bar{i}\bar{j}$ and the
index $\mu_3$ is a projection on the quantization axis z.
The double stretched quadrupole operators
$\tilde{Q}_{1}^{(-)}=\xi {\hat Q}_1^{(-)}$ ($\xi=\omega_x \omega_z/\omega_0^2$),
$\tilde{Q}_{2}^{(-)}= \eta {\hat Q}_2^{(-)}$ ($\eta=\omega_x \omega_y/\omega_0^2$)
are defined by means of the quadrupole operators
$\hat Q_m^{(r)}=i^{2 + m +(r+3)/2}( \hat Q_{2m}+(-1)^{(r+3)/2}\hat Q_{2 -m})/\sqrt{2(1+\delta _{m 0})}$,
where $\hat Q_{\lambda m}={\hat r}^\lambda Y_{\lambda m}$ (m=0,1,2).
The symmetry broken by the external rotational field (the cranking term)
implies
\begin{equation}
[\,H_{\Omega}\,,\,\hat J_y\mp i\hat J_z \,]\, = \pm\Omega (\hat J_y\mp i\hat J_z)
\label{spur}
\end{equation}
(hereafter, we use in all equations $\hbar=1$).
This condition is equivalent to the condition of the existence of the negative 
signature solution $\omega_\nu=\Omega$ created by the operator 
$\hat \Gamma^{\dagger}=(\hat J_z + i \hat J_y)/\sqrt{2 \langle \hat J_x \rangle}$ \cite{mar3}.
We recall that $\hat H_{\rm add}$ in $\hat H_0$ (Eq.(\ref{h1})) breaks Eq.(\ref{spur})
in general. However, to meet the condition (\ref{spur}) we determine the strength 
constant from the requirement of the existence of the RPA solution 
$\omega_\nu=\Omega$. As a result, the violation is unessential (see below).

We solve the RPA equations of motion for normal modes 
$[\hat H_{\Omega}, \hat O_{\nu}^\dagger]=\omega_{\nu}\hat O_{\nu}^\dagger$ with 
$\hat O_{\nu}^\dagger=\sum_\mu(\psi^{(\nu)}_{\mu}b^{+}_{\mu}-\phi^{(\nu)}_{\mu}b_{\mu})$ 
(cf \cite{JR}). The solution leads to a couple of equations for unknown coefficients
\begin{equation}
{\tilde R}^{\nu}_{1} = -\frac{1}{\sqrt{2}}\left[\hat O_\nu,\tilde Q_1^{(-)}\right],\qquad
{\tilde R}^{\nu}_{2} = \frac{i}{\sqrt{2}}\left[\hat O_\nu,\tilde Q_2^{(-)}\right]
\label{R12}
\end{equation}
Resolving these equations one obtains the secular equation
\begin{equation}
F(\omega_{\nu}) =  \det~({\bf D}- \frac{\bf 1}\chi)=0
\label{det}
\end{equation}
that determines all negative signature RPA solutions $\omega_{\nu}$. 
The matrix elements
$D_{km}(\omega_{\nu}) =\sum_\mu {\tilde f}_{k,\mu }
{\tilde f}_{m,\mu } C_\mu ^{km}/(E_\mu^2 -\omega_{\nu}^2)$
involve the coefficients $C_{\mu}^{km} = \omega_{\nu}$ for $k\neq m$ and
$E_{\mu }$ otherwise; ${\tilde f}_{m,\mu }$ are two-quasiparticle 
matrix elements of operators $\tilde Q_m^{(-)}$. Among collective solutions there 
are solutions that correspond to the shape fluctuations of the system and the 
rotational mode $\omega_\nu=\Omega$. With aid of Eq.(\ref{spur}) the system for 
the unknown coefficients ${\tilde R}^{\nu}_{1,2}$  can be cast in the 
form  similar to the classical expression for the wobbling mode 
\begin{equation}
\omega_{\nu=w} = \Omega \sqrt{\frac{[\Im_{x} - \Im^{\it eff}_{2}]
[\Im_{x} - \Im^{\it eff}_{3}]}{\Im^{\it eff}_{2}\Im^{\it eff}_3}}
\label{wob2}
\end{equation}
with microscopic  effective moments of inertia \cite{mar2}
\begin{equation}
\Im_{2,3}^{\it eff}=\Im_{y,z}+\Omega S\frac{\Im_x-\Im_{y,z}-\omega_\nu ^2S/\Omega}{\Im_{z,y}+\Omega S}
\label{mmoi}
\end{equation}
that depend on the RPA frequency. Here, $\Im_x =  \langle \hat J_x \rangle/\Omega$,
$S = \sum_\mu J^y_{\mu}J^z_{\mu}/(E^{2}_{\mu} -\omega_\nu ^2)$ and
$\Im_{y,z} = \sum_\mu E_{\mu}(J^{y,z}_\mu)^2/(E^{2}_{\mu} - \omega_\nu ^2)$.
Equation (\ref{wob2}) does not contain the solution $\omega_\nu=\Omega$. 

\begin{figure}[ht]
\includegraphics[height=0.3\textheight,clip]{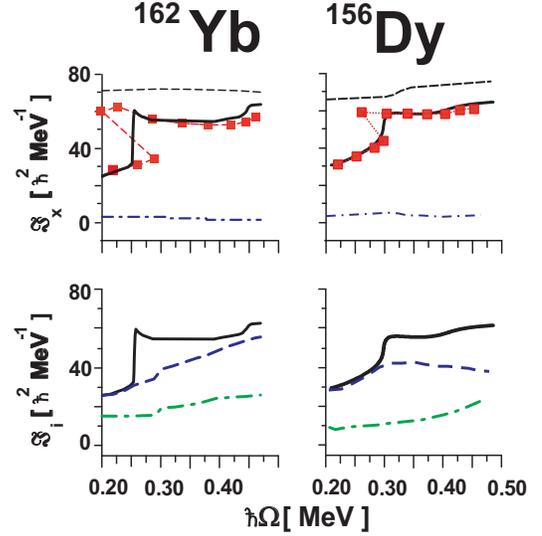}
\caption{(Color online)
Top panels: the kinematic
${\Im}_{x}= \langle{\hat J}_x\rangle/\Omega$
(solid line), the rigid body 
$\Im^{(rig)}_{1} = \frac{2}{5} m AR^{2} \left( 1- \sqrt{\frac{5}{4\pi}} \beta
\cos (\gamma - \frac{2\pi}{3}) \right)$ (dashed line)  and the hydrodinamical 
$\Im^{(irr)}_{1}= \frac{3}{2\pi}mAR^{2}\beta^{2}\sin ^{2} \left( \gamma -
\frac{2\pi}{3}\right)$ (dash-dotted  line)  moments of inertia
are compared with the experimental values (filled squares).
Experimental values ${\Im}_{x}= I/\Omega$ are connected by dashed line to guide eyes 
($\hbar\Omega=E_{\gamma}/2$).
Bottom panels display the rotational dependence of the kinematic moment of
inertia (solid line), effective moments of inertia $\Im^{\it eff}_{2}$ (dashed line) 
and $\Im^{\it eff}_{3}$ (dash-dotted line) for the first RPA solution $\nu=1$ obtained from
Eq.(\ref{det})}
\label{fig2}
\end{figure}
We obtain quite a remarkable correspondence between the experimental and calculated
values for the kinematic moment of inertia for both nuclei (see top panels in 
Fig.2).
The irrotational fluid moment of inertia $\Im^{(irr)}_1$
does not reproduce neither the rotational dependence nor the absolute values of the
experimental one as a function of equilibrium deformations (see Fig.2). 
The rigid body values provide the asymptotic limit of fast rotation without pairing,
if shell effects are smeared out (see discussion on shell effects at fast rotation 
in Ref.\onlinecite{del}).
Evidently, the difference between the rigid body and the calculated kinematic 
moments of inertia in both nuclei decreases with the increase of the rotational
frequency, although it remains visible at high spins.
At very fast rotation $\hbar\Omega>0.45$MeV the pairing correlations are
reduced due to multiple alignments, and, therefore, the difference is moderated.
It is evident that for the rotation around the axis $x$
the wobbling excitations with different collectivity could be found from
Eq.(\ref{wob2}), if $\Im_{x}>\Im^{\it eff}_{2},\Im^{\it eff}_{3} \quad (or
\quad \Im_{x}<\Im^{\it eff}_{2},\Im^{\it eff}_{3})$.
The rotational behavior of the effective moments of inertia for the first RPA solution
of Eq.(\ref{det}) (see Fig.2)  suggests that this solution may be associated with a wobbling mode.

To identify the wobbling mode among the solutions of Eq.(\ref{det}) 
it is instructive to introduce new variables, similar to ones in \cite{shi1} : 
$r_1^\nu={\tilde R_1^\nu}/(\xi A)$,  $r_2^\nu={\tilde R_2^\nu}/(\eta B)$,
where $A=\langle {\hat Q}_{2}+\sqrt{3}{\hat Q}_{0}\rangle$, $B=2\langle {\hat Q}_{2}\rangle$.
By means of Eqs.(\ref{R12}) and $\hat O_{\nu=\Omega}\equiv \hat\Gamma$,
we obtain exact definitions for the unknowns $r_{1,2}^\Omega$ associated with
the redundant mode $\omega_\nu=\Omega$: $r_1^\Omega=-1/2\sqrt{\langle \hat J_x \rangle}$, 
$r_2^\Omega=1/2\sqrt{\langle \hat J_x \rangle}$. With aid of these definitions,
exploiting the fact that the components of the quadrupole tensor commute, 
one can define the unknowns 
\begin{equation}
r_1^w=\frac{1}{2\sqrt{\langle \hat J_x\rangle}} \left(\frac{W_2}{W_3}\right)^{1/4}, \quad
r_2^w=\frac{1}{2\sqrt{\langle \hat J_x\rangle}} \left(\frac{W_3}{W_2}\right)^{1/4}
\label{wob4}
\end{equation}
and show (cf Ref.\onlinecite{shi1})  that they are associated with the wobbling 
mode. Here, $W_2=(1/\Im_2^{\it eff}-1/\Im_x)$, $W_3=(1/\Im_3^{\it eff}-1/\Im_x)$.
It is convenient to use the variables
$c_\nu=4\langle \hat J_x\rangle r_1^\nu r_2^\nu$.
From the definitions of $r_{1,2}^\Omega$, $r_{1,2}^w$ it follows that
\begin{equation}
c_{\nu=\Omega} \equiv -1, \quad c_{\nu=w} \equiv 1
\label{crit}
\end{equation}
Solving {\it only} the secular equation for the quadrupole
operators, Eq.(\ref{det}), the condition Eq.(\ref{crit}) enables us to
identify the redundant and the wobbling modes. Note that the variables
$r_{1,2}^\nu$ (or $c_\nu$) can be only defined for nonaxial shapes.

\begin{figure}[ht]
\includegraphics[height=0.34\textheight,clip]{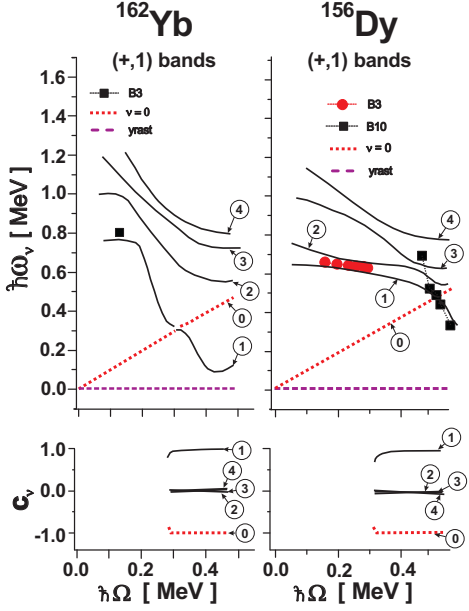}
\caption{(Color online)
Top panels: rotational dependence of the negative signature 
RPA solutions  with odd spins $(\pi=+,\alpha=1)$. The redundant mode 
$\omega_\nu=\Omega$ is denoted as "0" and is displayed by the dotted
line. Number in a circle denotes 
the RPA solution number : $1$ is the first $\nu=1$  RPA solution  etc.
Different symbols display the experimental data 
associated with B1,B2...bands (the band labels are taken in 
accordance with the definitions given in Ref.\onlinecite{nndc}).
 Bottom panels: the rotational dependence of the coefficients
$c_\nu \sim  r_{1}^\nu r_2^\nu$ (see text) that are determined
by the solutions of Eq.(\ref{det})}
\label{fig3}
\end{figure}
The experimental level sequences for all observed up-to-date 
rotational bands in $^{162}$Yb and $^{156}$Dy are
taken from Ref.\onlinecite{nndc}.
All rotational states are classified by 
quantum number $\alpha$ which is equivalent to our signature $r$.
The negative signature states ($r=-1$) correspond to $\alpha=1$ and are 
associated with odd spin states in even-even nuclei. All considered bands are 
of the positive parity $\pi=+$.
To elucidate the structure of  observed states,
 we define the experimental excitation energy in the rotating frame 
$\hbar \omega_{\nu}(\Omega)_{\it exp}=R_{\nu}(\Omega) - R_{yr}(\Omega)$ 
as a function of the rotational frequency $\Omega$ \cite{n87}. 
Here, the Routhian function 
$R_{\nu}(\Omega) = E_{\nu}(\Omega) - \hbar \Omega\, I_{\nu}(\Omega)$.
The energy $\hbar \omega_{\nu}(\Omega)_{\it exp}$ can be compared with the 
RPA results, $\hbar \omega_{\nu}(\Omega)$,  calculated 
at  a given rotational frequency. 

Top panels of Fig.3  display the redundant mode and four lowest RPA solutions of
Eq.(\ref{det}) as a function of the  rotational frequency.
We recall that these solutions are found at different equilibrium deformations
(see Fig.1). Indeed, in both nuclei the criteria Eq.(\ref{crit})
uniquely determines the redundant and the wobbling modes. In Fig.3
the redundant mode is manifested as a straight line (see  top panels),
while the corresponding coefficient $c_\Omega =-1$ (see bottom panels).
The redundant mode is separated clearly from the vibrational modes.

In $^{162}$Yb it is known only one negative signature $\gamma$-vibrational state.
The first RPA solution ($\nu=1$) is a negative signature gamma-vibrational
mode (with odd spins) till $\hbar\Omega\approx 0.28$MeV. With the increase of
the rotational frequency it is transformed to the wobbling
mode at $\hbar\Omega\approx 0.32$MeV (according to the criterion Eq.(\ref{crit})).
 Our results for $\nu=1$ solution may be used as a guideline for possible 
experiments on identification of the wobbling excitations near the yrast line. 
The first negative signature RPA solution in $^{156}$Dy  can be associated with the negative 
signature gamma-vibrational excitations with odd spins. After the transition from 
the axial to nonaxial rotation, at $\hbar\Omega\approx 0.3$MeV, according 
to the criteria Eq.(\ref{crit}), the first negative signature RPA solution describes 
the wobbling excitations.  The mode holds own features with the increase of 
the rotational frequency up to $\hbar\Omega \approx0.55$MeV. There is
a good agreement (see Fig.3, top right panel) between the RPA solution and  
the experimental  Routhian of band B10 (or $(+,1)_1$ band according to Ref.\onlinecite{Kon}). 
On this basis we propose to consider the B10 band as the wobbling band 
in the range of values $0.45$MeV$<\hbar \Omega < 0.55$MeV 
($33\,\hbar \leq I \leq 39\,\hbar$ for this band). 
Note that the band B10 contains  the states with $31\,\hbar - 53\,\hbar$. 
However, our conclusion is reliable only for the states 
with $I=33\hbar-\,39\,\hbar$ (or up to $\hbar \Omega < 0.55$MeV). 
At $\hbar \Omega \approx 0.55$MeV a crossing of the negative parity and 
negative signature (positive simplex) B6 band with the yrast band B8 is observed. 
Therefore, for $\hbar \Omega > 0.55$MeV (or for $I>39\,\hbar$ for the B10 band)
one may expect the onset of octupole deformation in the yrast states.
The octupole deformation is beyond  the scope of our analysis and  will be 
discussed in forthcoming paper.

In the microscopic approach \cite{shi1} the electric transition probabilities from 
the wobbling states  take the same form as in the macroscopic rotor model \cite{BM75}. 
Indeed, for interband transitions (from one-phonon to yrast states) we have 
(cf \cite{mar3,JR})
\begin{eqnarray}
\label{ntr}
&B(E2; I\, \nu\, \rightarrow \, I \pm 1 yr) \approx \\ 
&\left|\frac{i}{\sqrt{2}}\left[\tilde O_2^{(-)(E)},
\hat O_\nu^\dagger \right]/\eta \mp 
\frac{1}{\sqrt{2}}\left[\tilde O_1^{(-)(E)}, \hat O_\nu^\dagger \right]/\xi\right|^2
\nonumber
\end{eqnarray}
Here, $\hat{M}^{(E)}=(eZ/A) \hat{M}$.
In virtue of Eqs.(\ref{R12}),(\ref{wob4}), one can obtain   for
the quadrupole transitions from the one-phonon wobbling state to the
yrast states 
\begin{eqnarray}
\label{b1}
&B(E2; I\, w\, \rightarrow \, I \pm 1 yr) \approx \\
&\Biggl| \left(\frac{W_2}{W_3}\right)^{\frac{1}{4}}A^{(E)}
\mp \left( \frac{W_3}{W_2} \right)^{\frac{1}{4}}B^{(E)}\Biggr| ^2/(4\langle {\hat J}_x \rangle)
\nonumber
\end{eqnarray}
For intraband transitions we have (see \cite{mar3} and Eq.(43) in Ref.\onlinecite{JR})
\begin{equation}
B(E2; I \nu \rightarrow I - 2 \, \nu) \approx \frac{1}{8} \left|\sqrt{3}
\langle \hat{Q}^{(E)}_{0}\rangle -
\langle \hat{Q}^{(E)}_{2}\rangle \right|^{2}
\label{b2}
\end{equation}
Expressions (\ref{ntr}), (\ref{b1}), (\ref{b2}) are obtained in  high spin limit $I\gg1$.
To understand  a major trend of 
the quadrupole transitions, we employ relations from the pairing-plus-quadrupole 
model : $m\omega_0^2 \epsilon_2 \,cos\gamma^\prime = \chi \langle Q_{0}\rangle$,
$m\omega_0^2 \epsilon_2 \,cos\gamma^\prime = -\chi \langle Q_{2}\rangle$ 
(cf Ref.\onlinecite{fra}). By means of these relations
and  a definition of the quadrupole isoscalar strength 
$\chi= 4\pi m\omega^{2}_{0}/5\langle r^{2}\rangle \approx 4\pi m\omega^{2}_{0}/(3AR^{2})$
($R\approx1.2A^{\frac{1}{3}}fm$) one obtains from Eq.(\ref{b1})
\begin{eqnarray}
\label{bg}
&B(E2;\,I\,n_{w}=1 \rightarrow I\pm1 \, yr) \approx \\
&\Theta \epsilon_2^2
\Biggl[\Biggl(\frac{W_2}{W_3}\Biggl)^{\frac{1}{4}} sin(\frac{\pi}3-\gamma^\prime) \pm
\Biggl(\frac{W_3}{W_2}\Biggr)^{\frac{1}{4}} sin \gamma^\prime \Biggr]^2/{\langle \hat J_x\rangle},
\nonumber
\end{eqnarray}
where $\Theta=(9/16\pi^{2})e^{2}Z^{2}R^{4}$. 
Equation (\ref{bg}) yields the following selection rules for
the quadrupole transitions from the one-phonon wobbling band to the yrast 
one (for $W_{2,3} > 0$)
\begin{widetext}
\begin{eqnarray}
&&a) -60^{o}<\gamma<0 :\quad
B(E2; I\,n_{w}\rightarrow I - 1 yr) >
B(E2; I\,n_{w}\rightarrow I + 1 yr)\label{esr}\\
&&b)\quad 0<\gamma< 60^{o} :\quad
B(E2; I\,n_{w}\rightarrow I + 1 yr) >
B(E2; I\,n_{w}\rightarrow I - 1 yr)\nonumber
\end{eqnarray}
\end{widetext}

For the intraband transitions we obtain
\begin{equation}
B(E2;\,I\,n_{w}\rightarrow I-2 \, n_{w}) \approx \frac{1}{2}\Theta
\epsilon_2^2 cos^{2}(\frac{\pi}6-\gamma^\prime)
\label{tbyr}
\end{equation}
One observes from Eq.(\ref{tbyr}) that for the transitions along the yrast 
line ($n_w=0$) the onset of the positive (negative) values of $\gamma$-deformation 
leads to the increase (decrease) of the transition probability along the yrast line.
Moreover, the decay from one-phonon wobbling states to the yrast line
$R(\pm)=B(E2;\,I\,n_{w}=1 \rightarrow I\pm1 \, yr)/B(E2;\,I\,n_{w}\rightarrow I-2 \, n_{w}) \sim 1/I$
$(\langle \hat J_x\rangle\approx I \gg 1)$
decreases with the increase of the angular momentum for a constant deformation
$\gamma$. However, the rotational evolution of the nonaxiality may affect this
tendency. We predict almost a constant behaviour for the ratio $R(-)\approx 0.1$
for both nuclei at $\hbar\Omega>0.35$MeV due to the increase of the nonaxial deformation.

\begin{figure}[ht]
\includegraphics[height=0.3\textheight,clip]{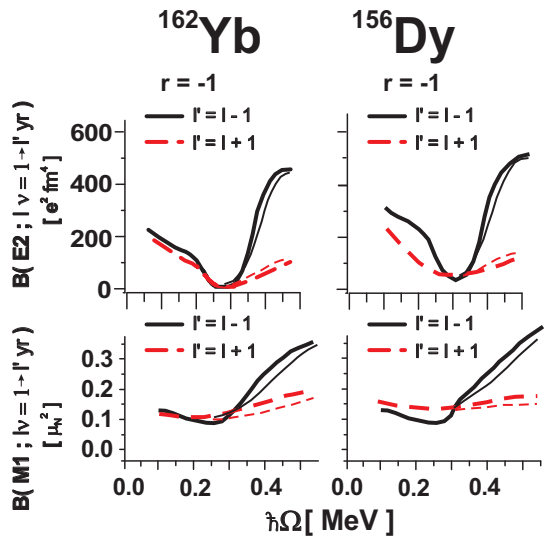}
\caption{(Color online)
B(E2)- (top) and  B(M1)- (bottom) reduced 
transition probabilities from the one-phonon bands to the yrast band. The 
negative signature phonon band is described by 
the first  RPA solution $(r=-1)$.
The transitions, calculated by means of the $\psi^{(\nu=1)}_{\mu}$ and 
$\phi^{(\nu=1)}_{\mu}$ phonon amplitudes,
are connected by solid lines. 
The results obtained 
by means of Eqs.(\ref{b1}), (\ref{mmt}) (with the aid of the variables $W_{2,3}$) 
are connected by thin lines, starting from the rotational frequency 
$\hbar\Omega\sim 0.3$ MeV. This point is associated in our analysis with 
the appearance of wobbling excitations.
One observes a strong dominance of the
B(E2)- and B(M1)-transitions from the wobbling states ($r=-1$) with spin I to
the yrast states with spin $I^\prime=I-1$ starting from the rotational
frequency $\hbar \Omega \geq 0.3$MeV.}
\label{fig4}
\end{figure}
At small rotational frequency, in  both nuclei, transitions probabilities 
from the first one-phonon states are much weaker 
than quadrupole transitions along the yrast line (compare with Fig.11 in Ref.\onlinecite{JR}).
At $\hbar\Omega \sim 0.05$MeV
the transition strength from the first one-phonon state to the yrast state : 
$\sim 330 e^2 fm^4 ( \sim 500 e^2fm^4)$
in $^{162}$Yb ($^{156}$Dy). We obtain a good 
correspondence between the shape evolution and the selection rules (\ref{esr}) 
for both nuclei (see top panels of Fig.4 and 
Fig.1). The transition probabilities, Eq.(\ref{ntr}), 
are calculated by means of the $\psi^{(\nu)}_\mu$ and $\phi^{(\nu)}_\mu$ phonon 
amplitudes. The results for the first negative signature RPA solution 
(which is associated with a wobbling mode) are compared  with those obtained with the aid of 
the effective moments of inertia (see Eqs.(\ref{mmoi}),(\ref{b1})). 
Evidently, if the "spurious" solution (the redundant mode) would be not removed 
from Eq.(\ref{det}), two estimations (\ref{ntr}) and (\ref{b1}) (based on different
secular equations (\ref{det}) and (\ref{wob2}), respectively) would produce  
different numerical values. A good agreement between  both 
results (see Fig.4) is the most valuable proof of the self-consistency of our calculations. 
The observed negligible differences are due to the approximate fulfillment of the conservation laws 
(\ref{spur}), caused by the additional term. In $^{162}$Yb, starting from 
$\hbar \Omega \sim 0.28$MeV (after the transition point), the  negative signature 
phonon band changes the decay properties. The interband quadrupole transitions 
from the one-phonon state  to the yrast ones with a lower spin 
dominate in the decay ($\Delta I=1$, the case Eq.(\ref{esr}), a)).
Similar results for the first negative signature one-phonon band 
are obtained in $^{156}$Dy. At low angular momenta 
($\hbar \Omega \leq 0.3$MeV) this band populates with approximately equal 
probabilities the yrast states with $I^\prime=I\pm1$ ($I$ is the angular momentum 
of the excited state). At $\hbar \Omega \sim 0.3$MeV a shape-phase transition 
occurs, that leads to the triaxial shapes with the negative $\gamma$-deformation. 
In turn, the phonon band
decays stronger on the yrast states with angular momenta $I^\prime=I-1$
($\Delta I=1$, the case Eq.(\ref{esr}), a)), starting from $\hbar\Omega\geq 0.32$MeV.

In the CRPA the magnetic transitions  are defined as (cf \cite{mar3,JR2}) 
\begin{equation}
B(M1; I\, \nu \rightarrow \, I\pm 1 \, yr) \approx
\frac{1}{2}\left|i \left[{\hat M}^{(M)}_{1\mu_{3}=1},{\hat O}_\nu^\dagger\right]
\mp \left[{\hat M}^{(M)}_{1\mu_{3}=0},{\hat O}_\nu^\dagger\right]\right|^2
\end{equation}
Here,  ${\hat M}^{(M)}_{1\mu_{3}}=
\mu_N\,\sqrt{3} \sum_{i=1}^{A}$  $($  $\frac{1}{2} g_s^{(i,eff)} [\sigma
\otimes Y_{l=0}]_{1\mu_{3}} \,+\,$
$g_l^{(i,eff)} [l \otimes Y_{l=0}]_{1\mu_{3}}$ $)$ is a magnetic dipole operator;
$\mu_{N}$ is the nucleon magnetons, $g_s^{(eff)}$, $g_l^{(eff)} $
are the spin and orbital effective gyromagnetic ratios, respectively.
Our results evidently demonstrate 
the dominance of $B(M1; I\,n_{W}\rightarrow I - 1 yr)$ (see  bottom 
panels in Fig.4) in both nuclei. In the rigid rotor model,
one can obtain for the magnetic transitions from the wobbling to yrast states
\begin{eqnarray}
\label{mmt}
B(M1; I\, \nu\, \rightarrow \, I\pm 1 \, yr) &&\approx
\frac{1}{4\langle {\hat J}_x \rangle}
\frac{(\sqrt{W_{3}}\mp\sqrt{W_{2}})^{2}}{\sqrt{W_{2}W_{3}}}\times\nonumber\\
&&\times \biggl|\langle \hat M^{(M)}_{1\nu_{3}=1}[r=+1]\rangle\biggr|^{2}
\end{eqnarray}
The full derivation will be presented elsewhere.
Note that the dipole magnetic moment $\langle \hat M^{(M)}_{1\nu_{3}=1}[+] \rangle$
increases quite drastically, if a nucleus is undergoing the backbending \cite{JR2}.
For  the wobbling states with $W_{2,3}>0$, Eq.(\ref{mmt}) yields 
\begin{equation}
B(M1; I\,n_{W}\rightarrow I - 1 yr) > B(M1; I\,n_{W} \rightarrow I + 1 yr)
\label{msr}
\end{equation}
At high spin limit $I\gg 1$, the microscopic and rigid body values of the variables
$W_{2,3}$ are very close. 
Thus, the macroscopic model supports the results of microscopic calculations for 
the magnetic transitions. It appears that the magnetic transitions with 
$\Delta I=1 \hbar$ always dominate from the wobbling to the yrast states,
independently from the sign of the $\gamma$-deformation 
of rotating nonaxial nuclei.  

In summary,  we predict that the lowest excited  negative signature and positive
parity band in $^{162}$Yb transforms to the wobbling band at
$\hbar\Omega\sim 0.3$MeV. We found that at $\hbar\Omega>0.25$MeV  
strong  E2-transitions from this band  start to populate yrast states, with the
branching ratio $B(E2; I\, w\, \rightarrow \, I - 1 yr)/B(E2; I\, w\, \rightarrow \, I +1 yr)>1$.
Similar transition occurs in $^{156}$Dy  after the backbending as well, at 
$\hbar\Omega>0.3$MeV.
A good agreement between our results  and experimental 
Routhians allows us to conclude that the experimental states, associated with 
$(+,1)_1$ band in $^{156}$Dy \cite{Kon}, are wobbling excitations at the rotational
frequency values $0.45$MeV$<\hbar\Omega<0.55$MeV. 
These states fulfill all requirements, specific for the wobbling excitations  of 
rotating triaxial nuclei with the negative $\gamma$-deformation.
It is quite desirable, however, to measure the interband $B(E2)$-transitions 
to draw a definite conclusion and we hope it will done in future.
We predict the dominance of $\Delta I=1 \hbar$ 
magnetic transitions from the wobbling to the yrast states, independently 
from the sign of the $\gamma$-deformation. 

\section*{Acknowledgments} 
We wish to thank the reviewer for his suggestions which helped to improve the 
clarity of the text and our analysis.
This work is a part of the research plan MSM
0021620859 supported by the Ministry of Education of the Czech Republic
and by the project 202/06/0363 of Czech Grant Agency.
It is also partly supported by Grant No. FIS2005-02796 (MEC, Spain).
R. G. N. gratefully acknowledges support from the
Ram\'on y Cajal programme (Spain).


\begin{thebibliography}{99}
\bibitem{prl}
                 E.\ S.\ Paul  {\it et al.},  
		 Phys.\  Rev.\  Lett.\ 98  (2007) 012501.
\bibitem{fra}
                 S. Frauendorf, 
                 Rev.\  Mod.\  Phys.\ 73 (2001) 463. 

\bibitem{BM75}
               A.\ Bohr and B.\ R.\ Mottelson,
               {\it Nuclear Structure} Vol. II
               (Benjamin, New York, 1975).

\bibitem{mic}
               I.\ N.\ Mikhailov  and  D. \ Janssen,
                Phys.\  Lett.\  B 72 (1978) 303;
               D.\ Janssen  {\em et al.},
               Phys.\  Lett.\  B 79 (1978) 347. 
	       
\bibitem{shi1}
               Y.\  R.\ Shimizu  and  M.\ Matsuzaki,
               Nucl.\  Phys.\ A 588 (1995) 559.

\bibitem{JM79}
               D.\ Janssen  and  I.\ N.\ Mikhailov,
               Nucl.\  Phys.\ A 318 (1979) 390.

\bibitem{mar2}
               E.\ R.\ Marshalek,
               Nucl.\  Phys.\ A 331 (1979) 429. 

\bibitem{OH01}
               S.\ W.\  O \O deg\aa rd  {\em et al.}, 
               Phys.\ Rev.\  Lett.\ 86 (2001 )5866;
               D.\ R.\ Jensen {\em et al.},  
               Phys.\  Rev.\  Lett.\  89 (2002) 142503;
               G.\ Sch\"{o}nwasser  {\em et al.}, 
               Phys.\  Lett.\ B 552 (2003) 9;
	       H.\ Amro  {\em et al.}, 
               Phys.\  Lett.\ B 553 (2003) 197;
               A.\ G\"{o}rgen {\em et al.},  
               Phys.\ Rev.\ C 69 (2004) 031301(R).
               
\bibitem{ham}
               I.\  Hamamoto,
               Phys.\ Rev.\ C 65 (2002) 044305.
              
\bibitem{mat}
               M.\ Matsuzaki,  Y.\ R.\ Shimizu, and K.\ Matsuyanagi,
               Phys.\ Rev.\ C 69 (2004) 034325. 

\bibitem{nndc}
               http://www.nndc.bnl.gov/nudat2/

\bibitem{Kon}  
               F.\ G.\ Kondev {\em et al.}, 
               Phys.\  Lett.\  B 437 (1998) 35. 

\bibitem{JR}
               J.\ Kvasil and R.\ G.\ Nazmitdinov,
               Phys.\  Rev.\ C 69 (2004) 031304(R);
               Phys.\  Rev.\ C 73 (2006) 014312.
\bibitem{nak}
               T.\ Nakatsukasa {\em et al.},
               Phys.\ Rev.\ C\ {\bf 53} (1996) 2213 .

\bibitem{KV}
               J.\ Kvasil  {\em et al.},
               Phys.\  Rev.\  C 58 (1998) 209.

\bibitem{wys}
               R.\ Wyss  {\em et al.}, 
               Nucl.\  Phys.\ A 511 (1990) 324. 

\bibitem{shel}
               A.\ K.\ Yain {\em et al.}, 
               Rev.\ Mod.\ Phys.\ 62 (1990) 393.
\bibitem{Nil}
               S.\ G.\ Nilsson {\em et al.},
	       Nucl.\ Phys.\ A131 (1969) 1.

\bibitem{mar3}
               E.\ R.\ Marshalek,
               Nucl.\  Phys.\ A  275 (1977) 416.
\bibitem{del}
               M.\ A.\ Deleplanque {\em et al.},
	       Phys.\ Rev.\ C 69 (2004) 044309.

\bibitem{n87}
               R.\ G.\ Nazmitdinov,
               Yad.\ Fiz.\  46 (1987)  732 
               [ Sov.\  J.\  Nucl.\  Phys.\ 46 (1987) 412 ].

\bibitem{JR2}
                J.\ Kvasil {\em et al.}, 
                Phys.\  Rev.\ C 69 (2004) 064308. 


\end{thebibliography}
\end{document}